\documentclass[pra,a4paper,twocolumn,superscriptaddress,longbibliography,nofootinbib]{revtex4-2}
\usepackage[utf8]{inputenc}
\usepackage{amssymb}
\usepackage{amsmath}
\usepackage{amsfonts}
\usepackage{amsthm}
\usepackage{amsbsy}
\usepackage{bm}
\usepackage{bbold}

\usepackage{graphicx}
\usepackage{xcolor}
\usepackage{multirow}
\usepackage{float}
\usepackage{comment}
\usepackage{ulem}
\usepackage{relsize}
\usepackage{cmap}
\usepackage[colorlinks]{hyperref}
\usepackage{fancyhdr}

\DeclareMathOperator{\sgn}{sgn}
\newcommand{\intl}{\int\limits}

\begin{document}

\title{Repulsion of a N\'eel-type skyrmion from a Pearl vortex in thin ferromagnet-superconductor heterostructures}

\author{E.\,S.\,Andriyakhina} 

\affiliation{\mbox{L. D. Landau Institute for Theoretical Physics, Semenova 1-a, 142432, Chernogolovka, Russia}} 

\affiliation{Moscow Institute for Physics and Technology, 141700 Moscow, Russia}

\author{S.\,Apostoloff}

\affiliation{\mbox{L. D. Landau Institute for Theoretical Physics, Semenova 1-a, 142432, Chernogolovka, Russia}}

\author{I.\,S.\, Burmistrov}

\affiliation{\mbox{L. D. Landau Institute for Theoretical Physics, Semenova 1-a, 142432, Chernogolovka, Russia}} 

\affiliation{Laboratory for Condensed Matter Physics, HSE University, 101000 Moscow, Russia}

\begin{abstract}
In this paper we study repulsion of a N\'eel-type skyrmion in a chiral ferromagnetic film from a superconducting Pearl vortex due to the stray fields. Taking into account an effect of the vortex magnetic field on the skyrmion non-perturbatively, we find that the repulsion between them is suppressed with increase of the dimensionless strength of the vortex magnetic field. This manifests itself  in complicated evolution of the free energy with increase of the vortex magnetic field and reduction of the equilibrium distance between the centers of N\'eel-type skyrmion and Pearl vortex. 
\end{abstract}

\maketitle

\thispagestyle{fancy}
\fancyhead{}
\fancyhead[LO,CE]{\textsf{\footnotesize submitted to JETP Letters}}
\fancyhead[RO,RE]{1}
\fancyfoot{} 

 Mutual influence of magnetism and superconductivity in heterostructures has long history of research~\cite{Ryazanov2004,Lyuksyutov2005,Buzdin2005,Bergeret2005,Eschrig2015}.
Recently, superconductor--ferromagnet (SF) bilayers hosting topologically nontrivial magnetic configurations have attracted much attention \cite{Back2020,Gobel2021,Zlotnikov}. Such topologically stable configurations can be stabilized by Dzyaloshinskii--Moriya interaction (DMI) in ferromagnetic films \cite{Bogdanov1989}. Skyrmions in SF heterostructures induce Yu-Shiba-Rusinov-type bound states \cite{Pershoguba2016,Poyhonen2016},
host Majorana modes \cite{Chen2015,Yang2016,Gungordu2018,Mascot2019,Rex2019,Garnier2019,Rex2020,Gungordu2022,Nothhelfer2022}, affect the Josephson effect \cite{Yokoyama2015}, and change the superconducting critical temperature \cite{Proshin2022}. 
 
Skyrmions and superconducting vortices  can form bound pairs in SF heterostructures due to interplay of spin-orbit coupling and proximity effect \cite{Hals2016,Baumard2019}. Also vortices and skyrmions interact via stray fields \cite{Dahir2019,Menezes2019,Dahir2020,Andriyakhina2021}. Recently, stable skyrmion-vortex coexistence has been experimentally observed in [Ir$_1$Fe$_{0.5}$Co$_{0.5}$Pt$_1$]$^{10}$/MgO/Nb sandwich structure \cite{Petrovic2021}.

In Ref. \cite{Andriyakhina2021} two of the authors predicted that a N\'eel--type skyrmion and a Pearl vortex interacting via stray fields are repelled from each other to be 
located at a finite distance.  However, the analysis of Ref. \cite{Andriyakhina2021}
has been limited to the lowest order perturbation theory in the magnetic field induced by the vortex. 

\begin{figure}[b]
\centerline{\includegraphics[width=0.9\columnwidth]{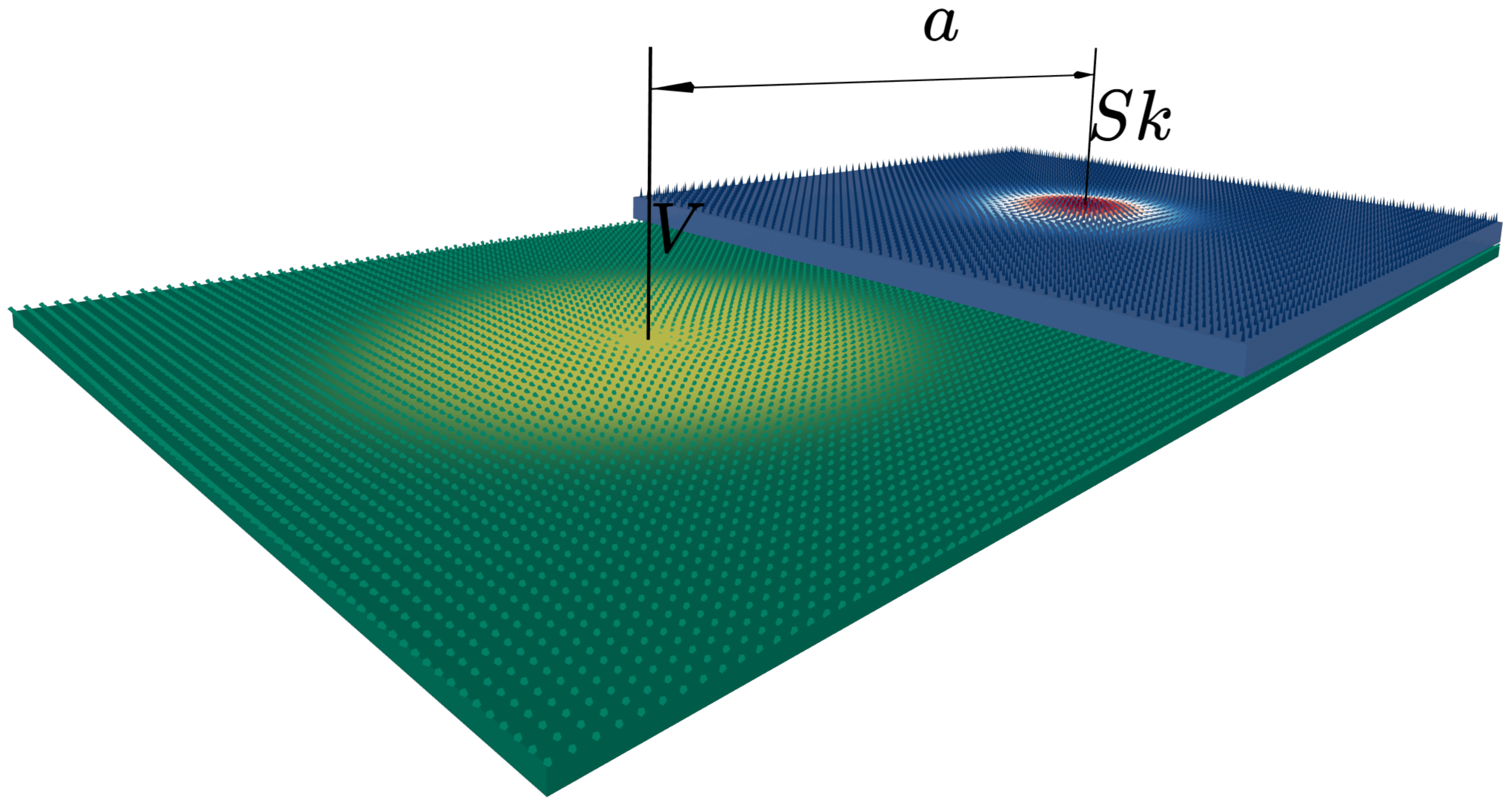}
}
\caption{Sketch of a ferromagnet (blue) -- superconductor (green) heterostructure. A thin insulating layer is not shown. The ferromagnetic layer hosts a N\'eel--type skyrmion (Sk). The superconducting film hosts a vortex (V). The distance between their centers is $a$.}
\label{Fig:setup}
\end{figure}

\begin{figure*}[t]
\centerline{\includegraphics[width=0.22\textwidth]{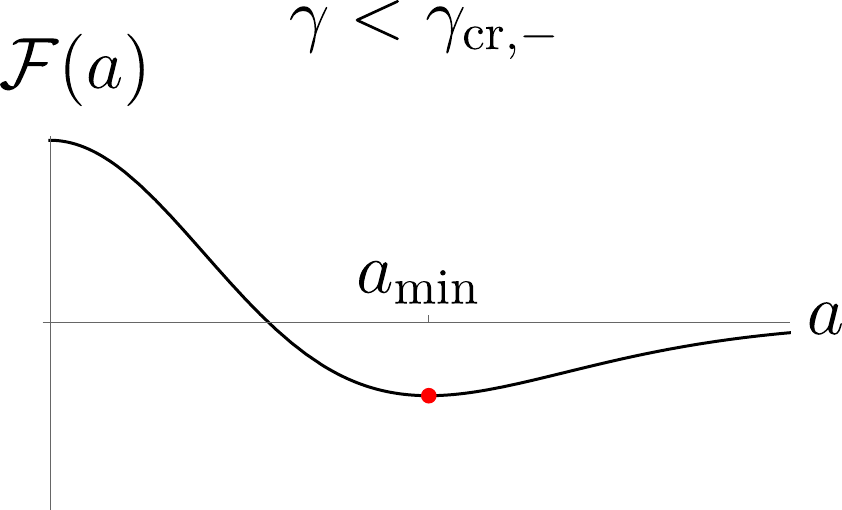}\quad\includegraphics[width=0.22\textwidth]{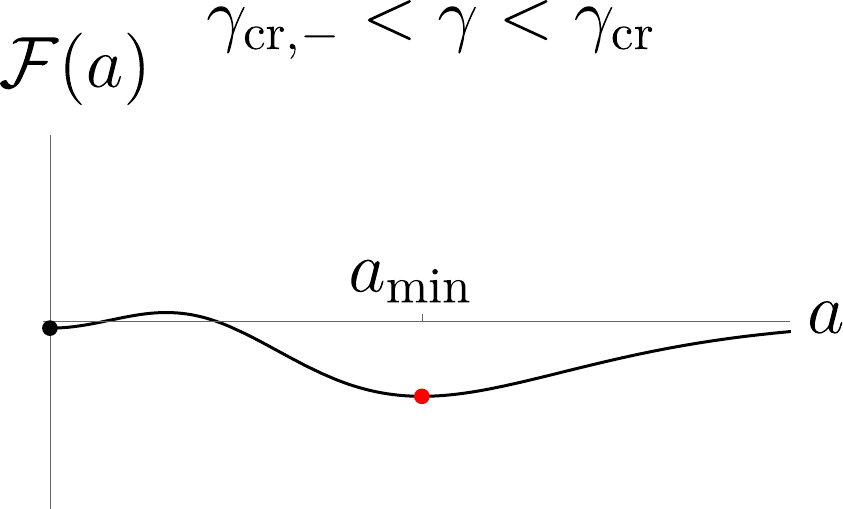}\quad
\includegraphics[width=0.22\textwidth]{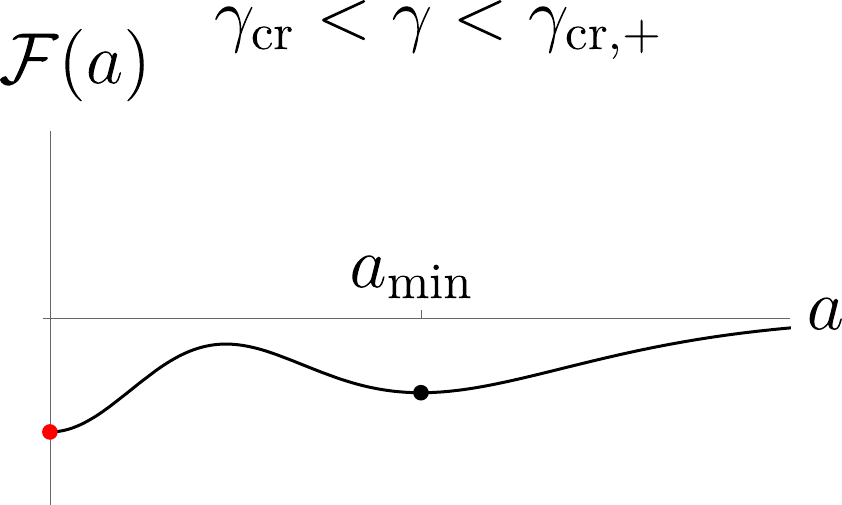}
\quad
\includegraphics[width=0.22\textwidth]{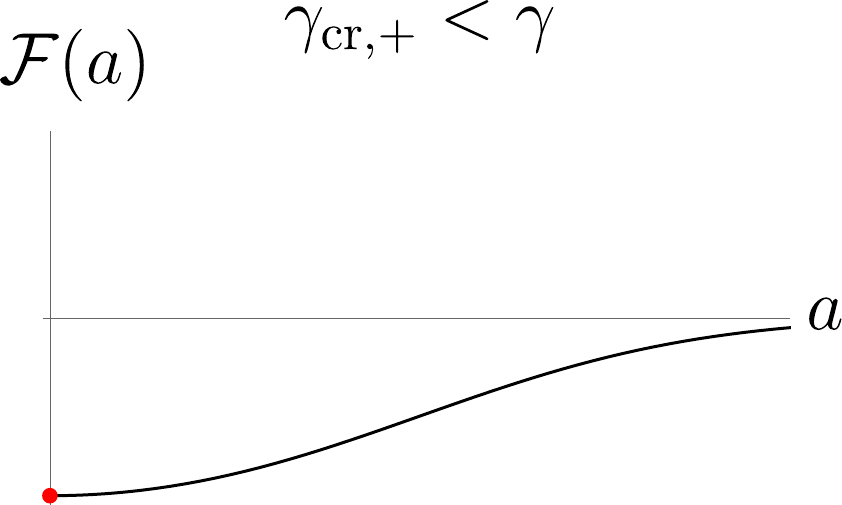}
}
\caption{Sketch of dependence $\mathcal{F}(a)$ for different ranges of $\gamma$. Left panel, $\gamma<\gamma_{\rm cr,-}$: There is the single minimum at nonzero distance $a_{\rm min}$. Middle panel with $\gamma_{\rm cr,-}<\gamma<\gamma_{\rm cr}$: There is the global minimum at nonzero distance $a_{\rm min}$ and the local one at $a=0$. Middle panel with $\gamma_{\rm cr}<\gamma<\gamma_{\rm cr,+}$: There is the global minimum at zero distance $a=0$ and the local one at $a_{\rm min}$. Right panel, $\gamma_{\rm cr,+}<\gamma$: There is the single minimum at $a=0$.}
\label{fig:FreeEnergy}
\end{figure*}

\begin{figure}[t]
\centerline{\includegraphics[width=0.9\columnwidth]{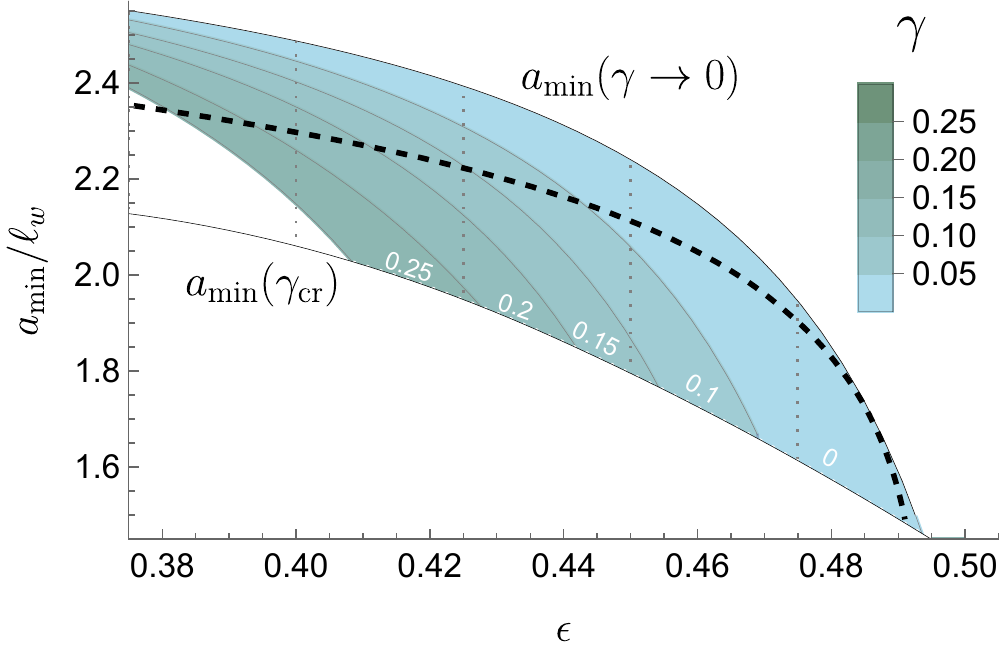}
}
\caption{Dependence of the distance $a_{\rm min}$ between the skyrmion and the vortex on $\epsilon$ for several values of $\gamma$ ranging from $0.01$ to $0.3$. Dashed curve illustrates analytical prediction for $a_{\rm min}(\epsilon)$ for $\gamma \to 0$, see text below Eq. \eqref{eq:b:asymp}. For each value of $\epsilon$ the distance $a_{\rm min}$ varies from $a_{\min}(\gamma \to 0)$ to $a_{\rm min}(\gamma_{\rm cr})$. Once $\gamma$ becomes larger than $\gamma_{\rm cr}$ the system reaches coaxial phase. The shaded regions guide the constant value of the vortex strength $\gamma$, see the color bar inset.}
\label{Fig:amin}
\end{figure}

\begin{figure}[t]
\centerline{\includegraphics[width=0.9\columnwidth]{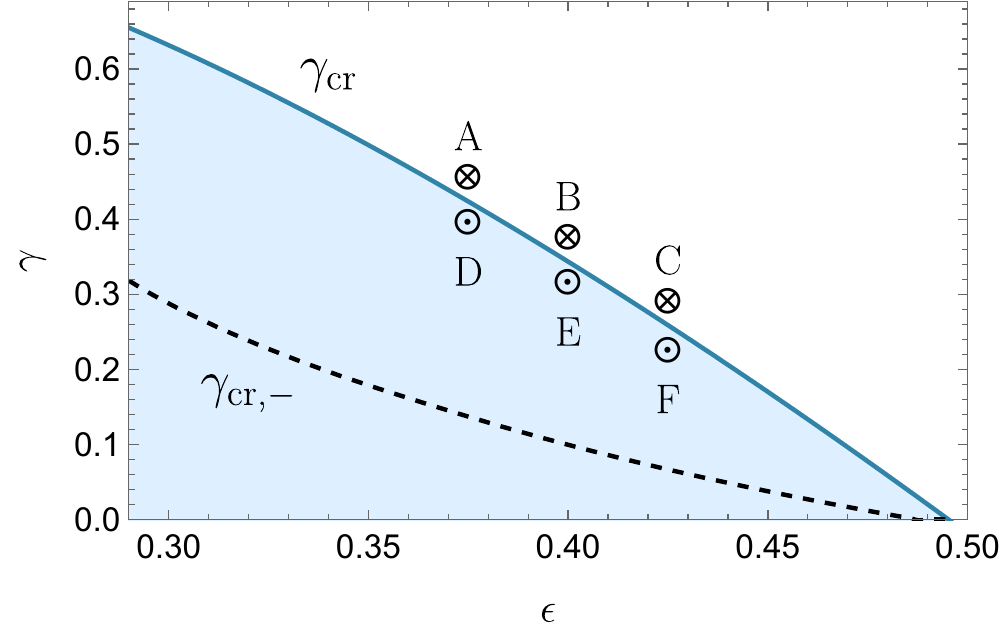}
}
\caption{Phase diagram. Solid curve shows dependence $\gamma_{\rm cr}(\epsilon)$ extracted from micromagnetic simulations. Black dashed curve corresponds to $\gamma_{\rm cr,-}(\epsilon)$. In blue shaded region the distance between the skyrmion and vortex is nonzero, $a_{\rm min}>0$, cf. Fig. \protect\ref{fig:FreeEnergy}. Points A, B, C, D, E, and F corresponds to the panels in Fig. \protect\ref{fig:SkV}. 
}
\label{Fig:Phases}
\end{figure}

In this Letter we study interaction between a superconducting Pearl vortex and a N\'eel-type skyrmion in a chiral ferromagnetic film due to stray fields (see Fig. \ref{Fig:setup}). Contrary to Ref. \cite{Andriyakhina2021}, we take into account the change of the skyrmion's profile due to magnetic field induced by the Pearl vortex. Performing non-perturbative treatment of the effect of vortex magnetic field on the one hand by analytical approach and on the other hand by micromagnetic modeling, we find that the free energy $\mathcal{F}$ of the system as a function of the distance $a$ between the skyrmion and the vortex experiences drastic changes with increase of the dimensionless strength~$\gamma$ of the vortex magnetic field, cf. Eq. \eqref{eq:gamma:def}. In particular, $\mathcal{F}$ 
 has 
 (i) the single minimum at $a=a_{\rm min}>0$ for $\gamma<\gamma_{\rm cr,-}$, 
 (ii) two minima at $a=0$ and at $a=a_{\rm min}$ with $\mathcal{F}(0)>\mathcal{F}(a_{\rm min})$ for $\gamma_{\rm cr,-}< \gamma<\gamma_{\rm cr}$, 
 (iii) two minima at $a=0$ and at $a=a_{\rm min}$ with $\mathcal{F}(0)<\mathcal{F}(a_{\rm min})$ for $\gamma_{\rm cr}< \gamma< \gamma_{\rm cr,+}$, and 
 (iv) the single minimum at $a=0$ for $\gamma_{\rm cr,+}<\gamma$ 
 (cf. Fig. \ref{fig:FreeEnergy}). With increase of $\gamma$ the distance $a_{\rm min}$ between the centers of a N\'eel-type skyrmion and a Pearl vortex is reduced and jumps to zero abruptly at $\gamma=\gamma_{\rm cr,+}$ (cf. Fig. \ref{Fig:amin}). All three critical values $\gamma_{\rm cr,\pm}$ and $\gamma_{\rm cr}$ depend on the dimensionless DMI strength (cf. Fig. \ref{Fig:Phases}). In general, we can make somewhat counter-intuitive statement: the repulsion between the skyrmion and the vortex is suppressed with increase of the dimensionless vortex magnetic field.

\noindent\textsf{\color{blue} Skyrmion-vortex interaction. --- }Following Ref. \cite{Andriyakhina2021}, our setup consists of ferromagnetic and superconducting films of thicknesses $d_F$ and $d_S$, respectively. We assume that both films are thin,
$d_S\ll\lambda_L$ and ${d_F\ll R}$,
where $\lambda_L$ is the London penetration length and $R$ denotes the skyrmion radius. Also we assume the presence of a thin insulating layer between superconducting and ferromagnetic films in order to suppresses the proximity effect. The superconducting film hosts a pair of Pearl vortex and antivortex separated by a distance much larger than 
the Pearl penetration length ${\lambda=\lambda_L^2/d_S}$ \cite{Pearl1964} 
(see Fig. \ref{Fig:setup}).

The free energy of a thin chiral ferromagnetic film subjected to the magnetic field ${\bm B}_{\rm V}$ produced by a Pearl vortex is given by
\begin{align}
	\mathcal{F}[\bm{m}] & =d_F \int d^2 \bm{r} \{ A (\nabla \bm{m})^2  + D [m_z \nabla {\cdot} \bm{m}- (\bm{m}{\cdot} \nabla) m_z ]
	\notag \\
	\, 
	&   + K(1- m_z^2) - M_s \bm{m}\cdot \bm{B}_{\rm V}|_{z=+0} \} .
	\label{eq:MagFe}
\end{align}
Here $\bm{m}(\bm{r})$ is the unit vector along direction of the magnetization $\bm{M}$, $M_s$ stands for  saturation magnetization of the film.
The exchange, 
DMI, and perpendicular anisotropy energy constants are denoted as $A$, $D$, and $K$, respectively. 
We assume that these parameters are positive, $A, K, D >0$. 
The magnetic field of the Pearl vortex centered at the position with coordinate $\bm{a}$ is given as \cite{AbrikosovBook}
\begin{eqnarray}
	{\bm B}_{\rm V}  = \phi_0 \sgn(z) \nabla 
	\int \frac{d^2\bm{q}}{(2\pi)^2} \frac{e^{-q |z| +i \bm{q}(\bm{r}-\bm{a})}}{q(1+2q\lambda)},
	\label{eq:vortex:B}
\end{eqnarray}
where $\phi_0=h c/2e$ denotes the flux quantum. 
The free energy~$\mathcal{F}[\bm{m}]$ is normalized in such a way that $\mathcal{F}=0$ for the ferromagnetic state, $m_z=1$, and without the Pearl vortex, ${\bm B}_{\rm V}=0$.
We note that we neglect interaction between the skyrmion and antivortex situated at large distance away.

\noindent\textsf{\color{blue} Shifted skyrmion at $\gamma\to0$. --- }In the absence of the vortex the magnetization corresponding to a N\'eel-type skyrmion can be sought in the cylindrical coordinate system as \cite{Kawaguchi2016}
\begin{equation}
    \bm{m} = \bm{e}_r \sin \theta (r)  + \bm{e}_z \cos \theta (r).
    \label{eq:m:profile}
\end{equation} 
Minimizing the free energy~$\mathcal{F}[\bm{m}]$ with ${\bm B}_{\rm V}=0$ by the skyrmion angle~$\theta(r)$, one can derive the Euler-Lagrange equation
, 
\begin{eqnarray}
	\ell_{w}^{2} 
	\Delta_r\theta(r)&-&\frac{(\ell_{w}^{2}+r^2)}{2 r^2}\sin2 \theta(r)
	+ 2\epsilon \frac{ \sin^2 \theta(r)}{r/\ell_{w}}  = 0 .
	\label{eq:ELE_theta_coax}
\end{eqnarray}
Here dimensionless parameter ${\epsilon=D/2\sqrt{AK}}$ controls the strength of DMI, the domain wall width ${\ell_{w} = \sqrt{A/K}}$ sets a natural length scale in the problem and  ${\Delta_r\theta=\partial_r(r \partial_r\theta)/r}$ is the radial part of Laplacian.

In order to solve Eq.~\eqref{eq:ELE_theta_coax} one needs to specify the boundary conditions. Assuming naturally that the film is ferromagnetically magnetized away from the skyrmion, $\theta(r\to \infty)=0$ ($m_z=1$). At the center of the skyrmion magnetization has the opposite direction, $m_z=-1$. The latter corresponds to the condition $\theta(r\to 0)=\pi$.

Knowing the solution $\theta(r)=\theta_{0}(r)$ of Eq.~\eqref{eq:ELE_theta_coax} (in the absence of vortex magnetic field), we can calculate~\cite{Andriyakhina2021} the interaction energy as function of distance $a$ between the centers of the skyrmion and the vortex,
\begin{equation}
    \dfrac{\delta\mathcal{F}(a)}{4\pi A d_F}=\gamma \int\limits_0^\infty dr \, r
	\{b_r^a(r)\sin\theta_0(r)+ b_z^a(r)[\cos\theta_0(r)-1]\}, 
	\label{eq:dFa1}
\end{equation}
where the dimensionless strength~$\gamma$ of the vortex magnetic field is given as 
\begin{equation}
    \gamma=(\ell_{w}/\lambda)(M_s\phi_0/8\pi A) .
\label{eq:gamma:def}
\end{equation}
Equation \eqref{eq:dFa1} is valid in the leading approximation on small vortex strength $\gamma\ll1$. Notation $\delta\mathcal{F}$ means that we subtract the energy of the lonely skyrmion itself and the energy of the homogeneous ferromagnet in the vortex field from the total free energy $\mathcal{F}[\bm{m}]$. The functions $b_r^a(r)$ and $b_z^a(r)$ means the dimensionless $r$- and $z$-projections of the vortex field ${\bm B}_{\rm V}$ averaged by rotation of the system around the center of the skyrmion, i.e. over all possible directions of the vector $\bm{a}$. They are given as 
\begin{gather} 
	 b_{r/z}^a(r)=2\ell_{w}\intl_{0}^{\infty} \frac{dq\,qJ_0(qa)J_{1/0}(qr) }{(\lambda^{-1}+2q)}.
\label{eq:b:def}
\end{gather}
Assuming skyrmion radius to be much smaller than the Pearl length, $R\ll \lambda$, that requires $d_S\ll\lambda_L^2/R$, we can treat functions $b_{r/z}^a(r)$ in the limit  $r\sim R\ll \lambda$ as
\begin{equation}
    \begin{split}
	b_{z}^a(r) \approx\frac{2\ell_w}{\pi  (a+r)}K\Big(\frac{4 a r}{(a+r)^2}\Big ),
	\\
	b_{r}^a(r) \approx (\ell_w/r)\Theta(r-a) .
\end{split}
\label{eq:b:asymp}
\end{equation}
Here $\Theta(z)$ denotes the Heaviside step function and $K(z)$ is the complete elliptic integral of the first kind.

The minimum of $\delta\mathcal{F}(a)$ determines the stable position $a_{\rm min}$ of the skyrmion. The resulting dependence of $a_{\rm min}$ on $\epsilon$ at $\gamma\to 0$ is shown in Fig. \ref{Fig:amin} by dashed line. As one can see, $a_{\rm min}$ is of the order of $2\ell_w$ and decreases with increase of $\epsilon$. 

It should be emphasized that in the analysis above we neglect a change of the skyrmion profile due to the vortex field when calculating Eq.~\eqref{eq:dFa1}. Indeed, such a reshaping should be taken into account in the next approximation on small vortex strength $\gamma$.
For small $\gamma$ an account of the next order correction leads only to a small correction of the magnitude of $a_{\rm min}$, but does not change the result qualitatively.

\noindent\textsf{\color{blue} Nearly centered skyrmion, $a{\to}0$. --- }Now we study at what parameters the coaxial configuration of the skyrmion and the vortex ($a=0$) is unstable. The magnetization of such skyrmion also can be sought as given by Eq. \eqref{eq:m:profile}, 
due to the radial symmetry of the problem. Then the minimization of the free energy~$\mathcal{F}[\bm{m}]$ by the skyrmion angle~$\theta(r)$ yields the Euler-Lagrange equation similar to Eq.~\eqref{eq:ELE_theta_coax}, but with $\gamma [b_{r}^0(r)\cos\theta(r)-b_z^0(r)\sin\theta(r)]$ instead of zero in the right hand side. Boundary conditions remain the same: $\theta(r\to \infty)=0$ and $\theta(r\to \infty)=\pi$.

When the solution $\theta(r)=\theta_\gamma(r)$ for a finite magnitude of $\gamma$ is obtained, we can calculate the interaction energy~$\delta\mathcal{F}(a)$ for small $a\ll\ell_w$. The idea is the same as beyond Eq.~\eqref{eq:dFa1}. For small spacing $a$, one can neglect the reshaping of the skyrmion in the leading approximation and use Eq.~\eqref{eq:dFa1}, but with $\theta_\gamma(r)$,
\begin{eqnarray}
    \dfrac{\delta\mathcal{F}(a)-\delta\mathcal{F}(0)}{4\pi A d_F}=\gamma \int\limits_0^\infty dr \, r
	\{&& \delta b_r^a(r) \sin\theta_\gamma(r)
	\notag\\
	&&{}+ \delta b_z^a(r)[\cos\theta_\gamma(r)-1]\}. 
	\label{eq:dFa2}
\end{eqnarray}
Here we introduce $\delta b_{r/z}^a(r)= b_{r/z}^a(r)-b_{r/z}^0(r)$. At 
$r\sim R\ll \lambda$ and $a\to 0$ these functions can be estimated as $\delta b_{z}^a(r)\approx \ell_w a^2/(4 r^3)$ and $\delta b_{r}^a(r)\approx \ell_w a^2/(8 r^2\lambda)$. Note that though the result~\eqref{eq:dFa2} has been obtained in the limit of small $a$, it allows to predict the existence of a minimum of $\mathcal{F}(a)$ at $a=0$ for an arbitrary value~$\gamma>0$.

At $\gamma\to 0$ and for $\epsilon<\epsilon_{\rm cr,-}\approx 0.488$ the free energy has a maximum at $a=0$, because $\delta\mathcal{F}(a)<\delta\mathcal{F}(0)$. Hence, the skyrmion is repelled from the vortex in agreement with Ref. \cite{Andriyakhina2021}. With increase of $\gamma$ the difference $\delta\mathcal{F}(a)-\delta\mathcal{F}(0)$ 
changes sign at $\gamma=\gamma_{\rm cr,-}$ and the minimum at $a=0$ appears in 
the free energy. Therefore, for $\gamma>\gamma_{\rm cr,-}$ the skyrmion can be stable right at the vortex. The dependence of $\gamma_{\rm cr,-}$ on $\epsilon$ extracted from Eq. \eqref{eq:dFa2} is shown in Fig. \ref{Fig:Phases} by the dashed curve.  As one can see $\gamma_{\rm cr,-}$ decreases with increase of $\epsilon$ and vanishes at ${\epsilon_{\rm cr,-}\approx 0.488}$. We expect similar dependence of $\gamma_{\rm cr,+}$ on $\epsilon$. Our theoretical analysis demonstrates that $\gamma_{\rm cr,+}$  vanishes at $\epsilon_{\rm cr,+}\approx0.493$ which is only slightly larger than $\epsilon_{\rm cr,-}$. We note that  
$\gamma_{\rm cr}$ vanishes at the value of $\epsilon_{\rm cr}\approx0.491$ which lies between $\epsilon_{\rm cr,-}$ and $\epsilon_{\rm cr,+}$. We stress that with relevant precision $\epsilon_{\rm cr,-} \simeq \epsilon_{\rm cr}\simeq \epsilon_{\rm cr,+}\approx 0.49$.

\begin{figure*}[t]
    \centering
    \begin{tabular}{lll}
        (A): $\epsilon = 0.375$, $\gamma = 0.43$ & (B): $\epsilon = 0.4$, $\gamma = 0.35$ & (C): $\epsilon = 0.425$, $\gamma = 0.26$ \\
        \includegraphics[width=.3\textwidth]{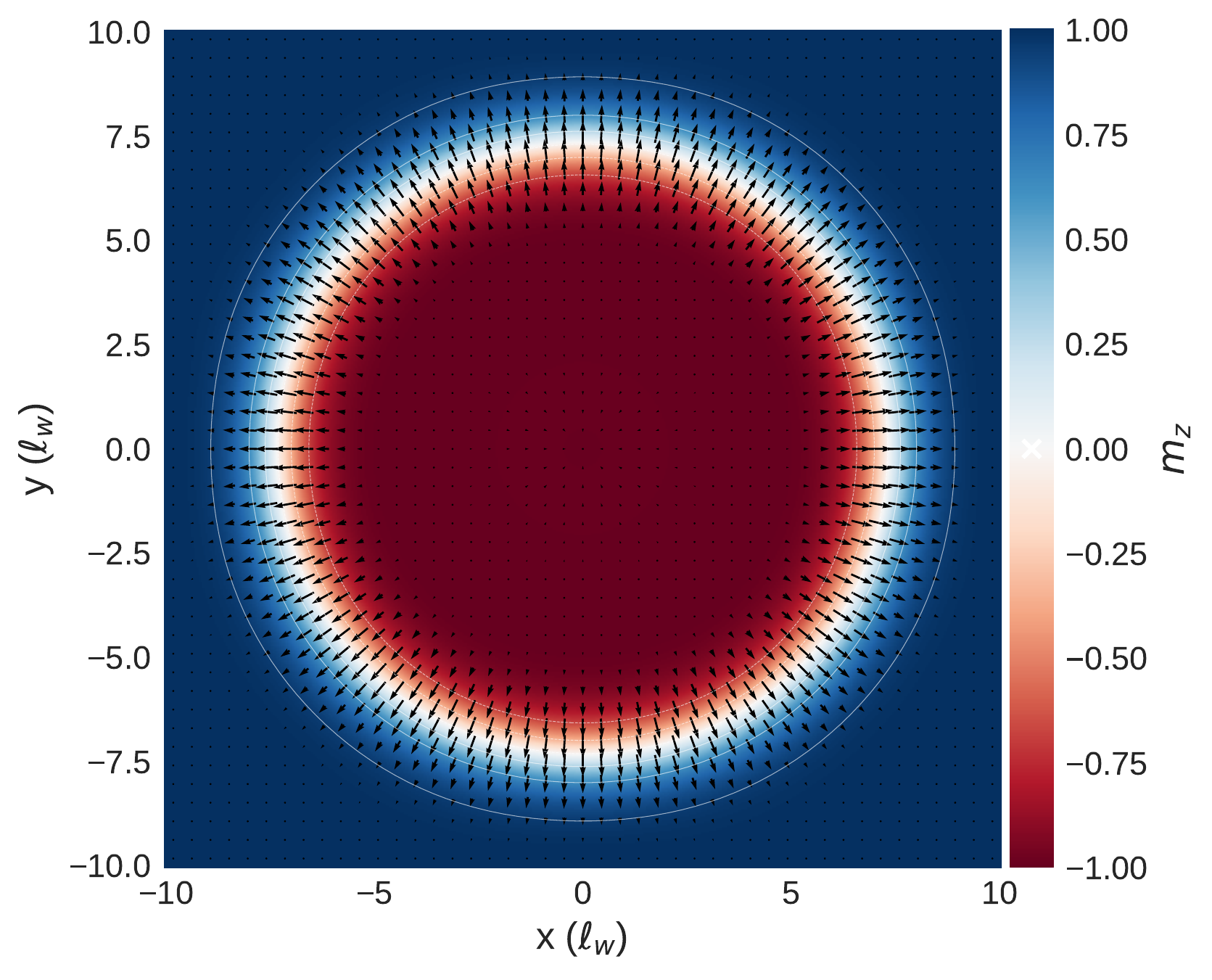} &
        \includegraphics[width=.3\textwidth]{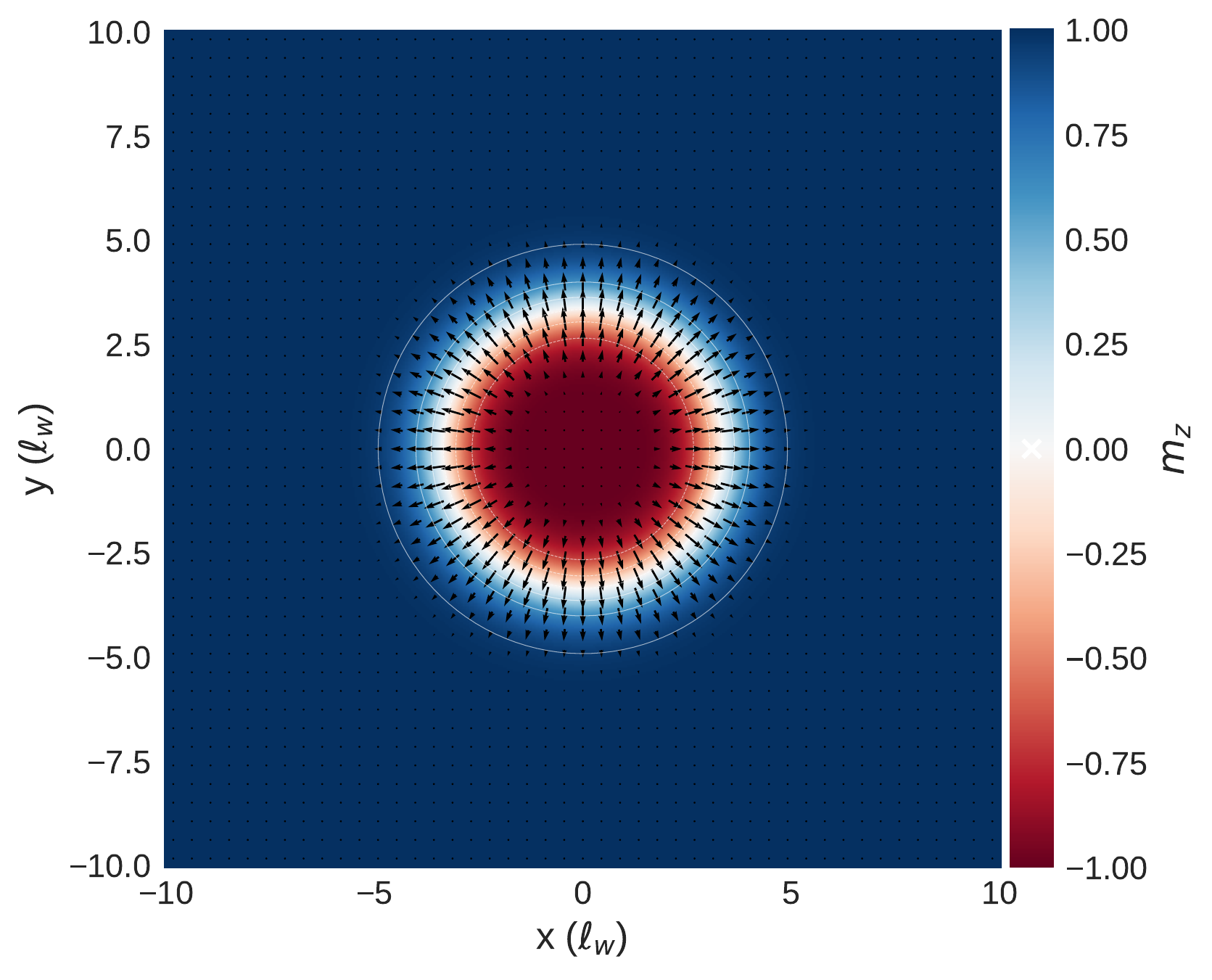} & \includegraphics[width=.3\textwidth]{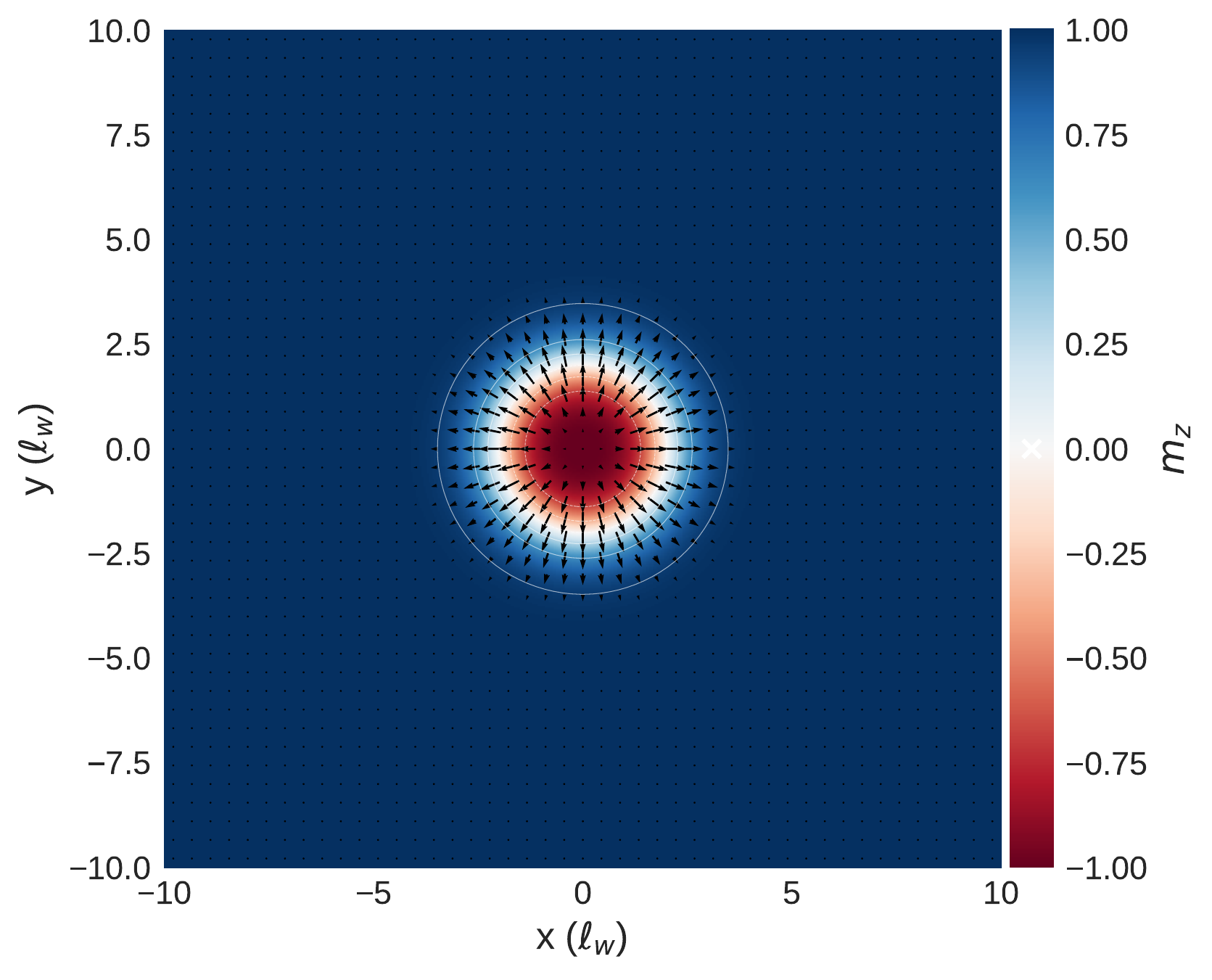} \\
        (D): $\epsilon = 0.375$, $\gamma = 0.42$ & (E): $\epsilon = 0.4$, $\gamma = 0.34$ & (F): $\epsilon = 0.425$, $\gamma = 0.25$ \\
        \includegraphics[width=.3\textwidth]{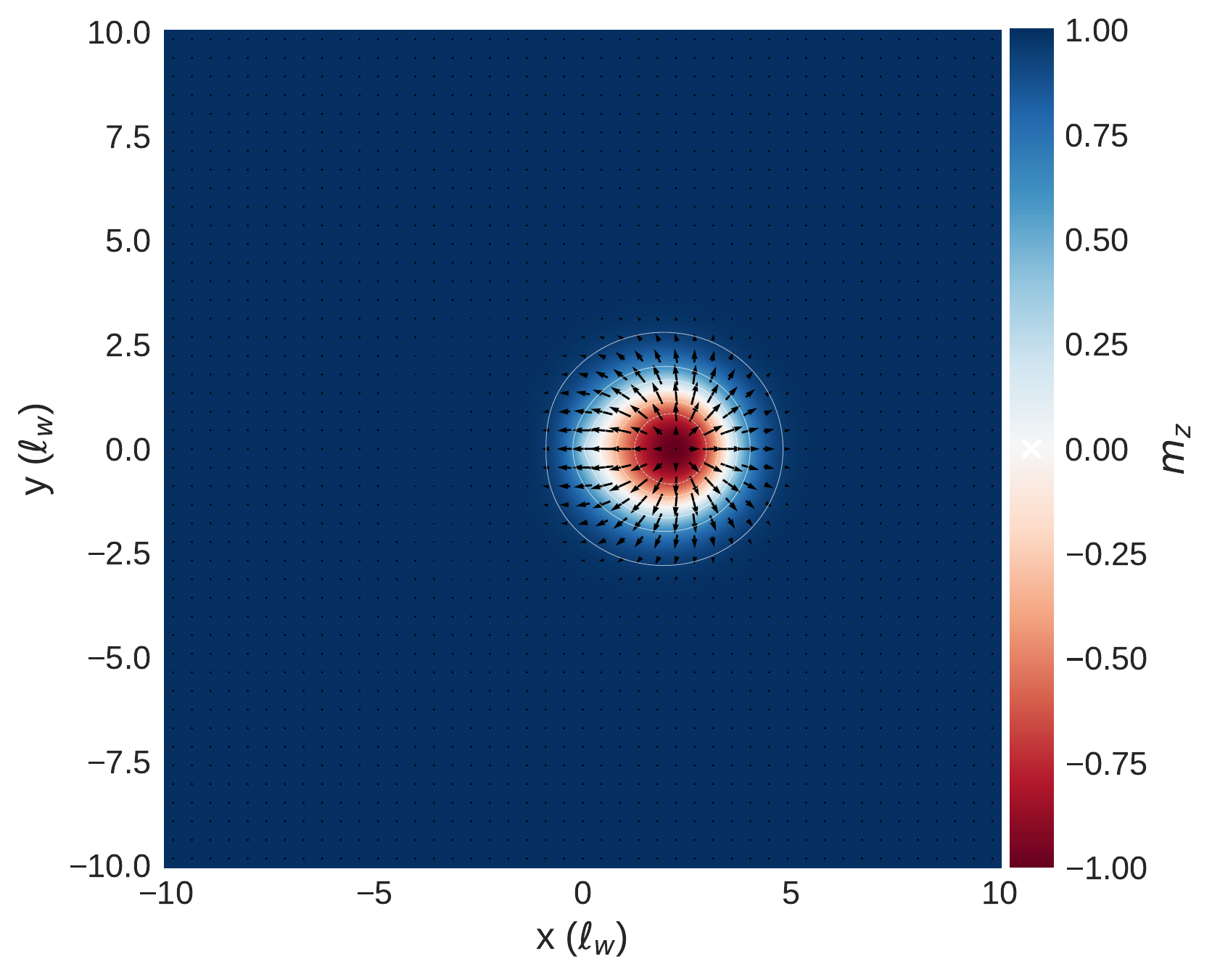} &
        \includegraphics[width=.3\textwidth]{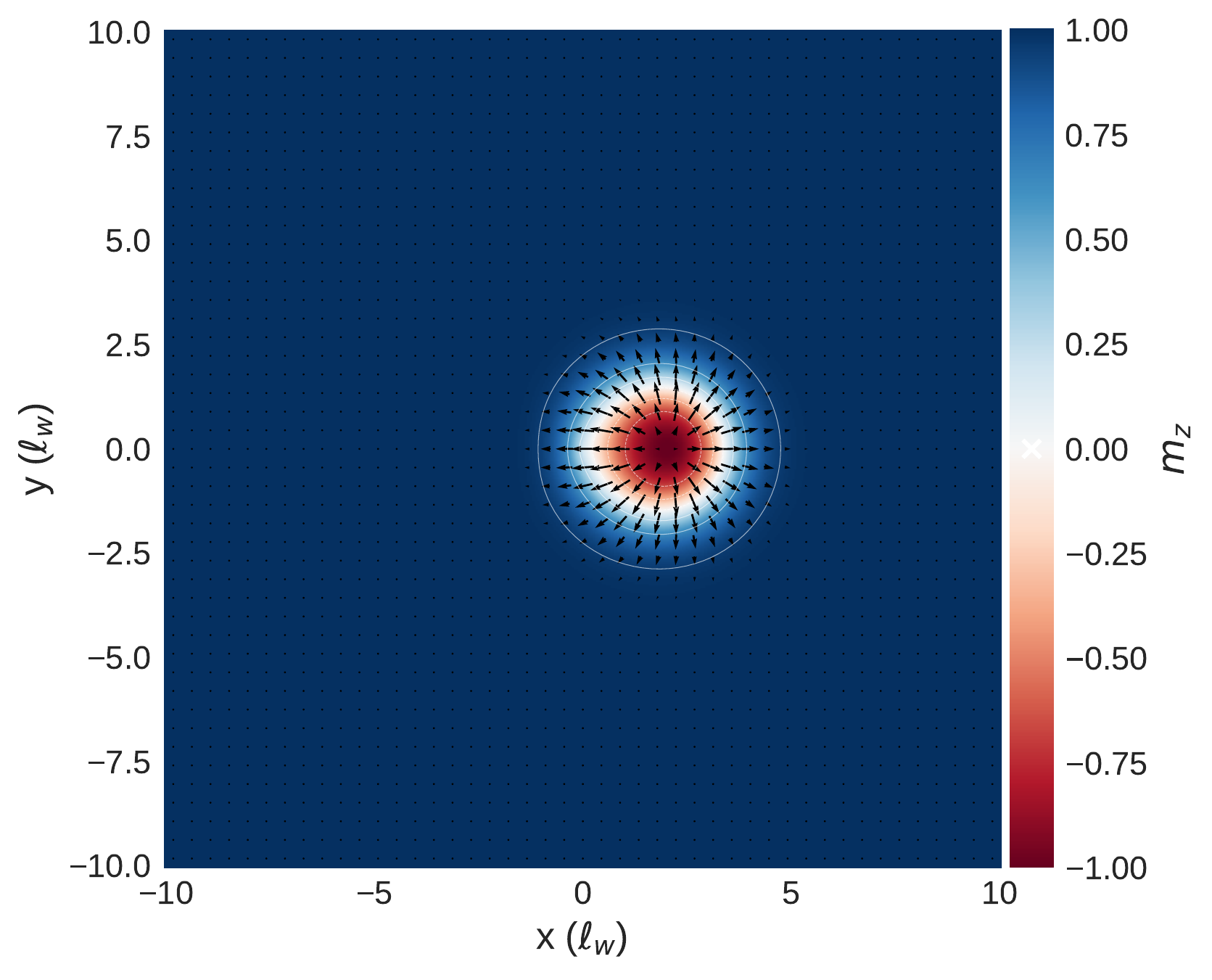} & \includegraphics[width=.3\textwidth]{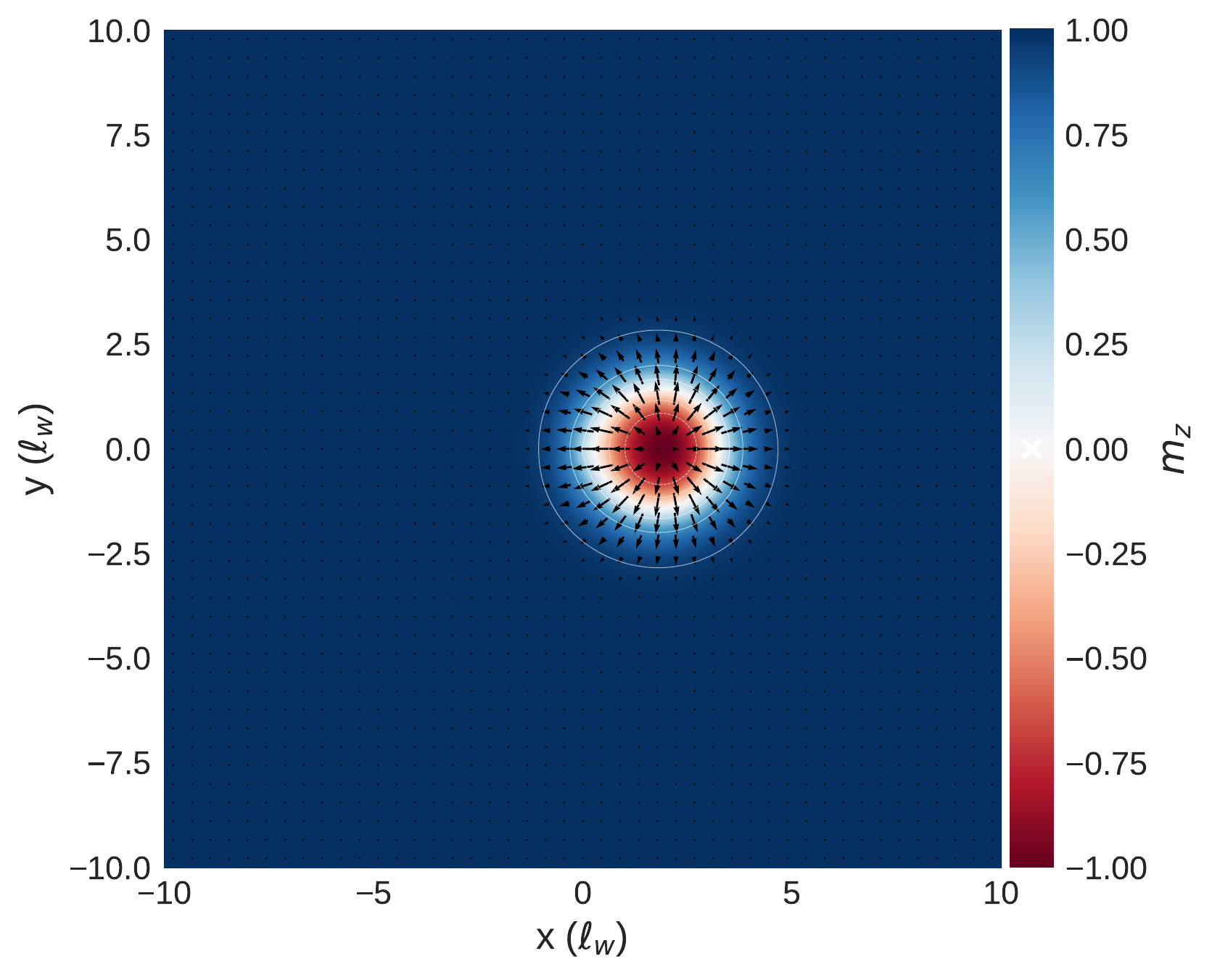}
        \end{tabular}
    \caption{Distributions of magnetization for different values of $\epsilon$ and  $\gamma$. The Pearl vortex is located at the center of each picture, i.e. at $x = y = 0$. White contours illustrate 
   a level set of the $m_z$ projection. The most extreme contour in each panel separates the skyrmion magnetization from purely ferromagnetic spin ordering. Upper row corresponds to vortex strength $\gamma$ slightly above the critical $\gamma_{\rm cr}$, when  skyrmion abruptly changes its position and becomes coaxial with the vortex. Lower row illustrates distribution of 
   $m_z$ for $\gamma$ which is slightly less than $\gamma_{\rm cr}$, i.e. for the case of nonzero $a_{\rm min}$. One can clearly 
   see 
  that skyrmions in the lower row are shifted from the sample's center where the Pearl vortex is settled. The panels (A), (B), (C), (D), (E), and (F) match with the points marked in Fig. \protect\ref{Fig:Phases}. \color{black}}
    \label{fig:SkV}
\end{figure*}

\noindent\textsf{\color{blue} 
Micromagnetic modeling. --- } To explore the skyrmion stable state in the field of the Pearl vortex with change of $\epsilon$ and $\gamma$, we perform micromagnetic simulations. 
We utilize Object Oriented MicroMagnetic Framework (OOMMF) \cite{OOMMF} by dint of Ubermag \cite{Ubermag} Python packages. 

The system is modeled as a set of classical magnetic vectors placed in the center of mesh cells. The distance 
is measured in units of the domain wall width $\ell_{w}$. 
We note that for micromagnetic modeling we set the magnetic anisotropy parameter $K=1$ and DMI $D=2\epsilon$.
Periodical boundary conditions (in $xy$ - plane) are imposed to simulate an isolated region of a ferromagnet-insulator-superconductor structure. We create a skyrmion by initiating a closed region with a flipped magnetization and letting it relax in the presence of Heisenberg exchange, DMI and magnetic anisotropy interactions as well as in the vortex-induced magnetic field. 

We point out that the system is not subject to any other external fields, so that the only source of Zeeman energy comes from the interaction with the vortex. In our simulations we consider a Pearl vortex with zero-sized core to be pinned in the origin of the grid, see Eq.~\eqref{eq:vortex:B}. 
The micromagnetic simulations allow us to find dependence of the distance~$a_{\rm min}$ between the skyrmion and the vortex on~$\epsilon$ for a finite value of~$\gamma$, see Fig.~\ref{Fig:amin}. As in the case of $\gamma\to 0$, $a_{\rm min}$ decreases with increase of $\epsilon$ at constant magnitude of $\gamma$. Similarly, $a_{\rm min}$ decreases with increase of $\gamma$ at constant value of $\epsilon$. We note a discrepancy between the theory and the micromagnetic simulations that becomes more prominent for smaller $\epsilon$, see Fig. \ref{Fig:amin}. We believe that this happens due to discretization effects which are inevitable in the numerical approach. In our simulations we have found that by making the mesh finer, one observes the declining tendency for $a_{\rm min}$ towards the theoretical value marked by the dashed curve in Fig. \ref{Fig:amin}. However, for small values of $\epsilon$ one needs to account both for smaller mesh cells and for larger sample size, which is computationally expensive. 

The results of micromagnetic simulations are consistent with evolution of the free energy with $\gamma$ illustrated in Fig.~\ref{fig:FreeEnergy} schematically. The phase diagram in $\epsilon$ and $\gamma$ plane extracted from the obtained results is shown in Fig.~\ref{Fig:Phases}. Two phases of stable position of the skyrmion, right on the top of the vortex (white clear upper region) and at the finite distance $a_{\rm min}$ (blue shaded lower region), are separated by the solid line $\gamma=\gamma_{\rm cr}(\epsilon)$. One can see that $\gamma_{\rm cr}$ drops to zero when $\epsilon$ reaches approximately $0.49$ in accordance with theoretical predictions. As we have mentioned above there is the lower and upper critical values $\gamma_{\rm cr,\mp}(\epsilon)$ at which minimum at $a=a_{\rm min}>0$ disappears and minimum at $a=0$ appears, respectively. However, we cannot resolve these values within our micromagnetic modeling. 

The skyrmion's profiles obtained by means of micromagnetic simulations for $\epsilon$ and $\gamma$ corresponding to the points A, B, and C in Fig.~\ref{Fig:Phases} are presented in Fig.~\ref{fig:SkV} (upper row). As expected, in all three cases skyrmions are situated right on the top of the vortex, i.e. $a_{\rm min}=0$. In lower row of Fig.~\ref{fig:SkV} we demonstrate skyrmions for parameters $\epsilon$ and $\gamma$ corresponding to the points D, E, and F in Fig.~\ref{Fig:Phases}. In this case the nonzero distance of the skyrmion from the vortex is clearly seen. 

To clarify significance of critical values of dimensionless vortex strength $\gamma_{\rm cr,\pm}$ and $\gamma_{\rm cr}$ let us consider the case of small but finite concentration of skyrmions and vortices. Then at $\gamma<\gamma_{\rm cr,-}$ one can expect the phase of (dipole-like) pairs consisting of skyrmion and vortex separated by a distance $a_{\rm min}$. At $\gamma>\gamma_{\rm cr,+}$ one can observe the phase of (point-like) pairs of skyrmion and vortex sitting on the top of each other. In the intermediate range $\gamma_{\rm cr,-}<\gamma<\gamma_{\rm cr,+}$ there is the phase in which there are finite concentrations of dipole-like and point-like skyrmion-vortex pairs. Under assumption that the system can reach 
the global minimum of the free energy, there will be a true thermodynamic  transition at $\gamma_{\rm cr}$ between the phases with dipole-like and point-like pairs, respectively. Since the dimensionless vortex strength $\gamma$ depends on material parameters, see Eq. \eqref{eq:gamma:def}, and is proportional to the thickness of superconducting film $d_S$, it could be possible to see the transitions described above with thickness change. 

\noindent\textsf{\color{blue} Summary. --- } In this paper we extended the study of interaction between a superconducting Pearl vortex and a N\'eel-type skyrmion in a chiral ferromagnetic film to non-perturbative regime with respect to the stray fields induced by the vortex. Contrary to the previous work \cite{Andriyakhina2021} of two of us limited to the regime of a weak dimensionless vortex magnetic field, $\gamma\to0$, we found that the increase of $\gamma$ suppresses repulsion of skyrmion and vortex and leads to reduction of the distance $a_{\rm min}$ between the centers of a N\'eel-type skyrmion and a Pearl vortex as shown in Fig.~\ref{Fig:amin}. Most surprisingly, we discovered the existence of interesting evolution of the free energy of the system with $\gamma$. In particular, at $\gamma<\gamma_{\rm cr,-}$ the free energy $\mathcal{F}(a)$ has the only minimum at  $a=a_{\rm min}$ whereas at $\gamma>\gamma_{\rm cr,+}$ it has the only minimum at $a=0$, see Figs.~\ref{fig:FreeEnergy} and~\ref{Fig:Phases}. 

Finally, we mention that it would be interesting to generalize our results to the case of skyrmions and vortices in confined geometries, e.g. nanodots etc.  
\cite{Rohart2013,Vadimov2018,Gonzalez2022}, 
skyrmion-vortex lattices \cite{Neto2022}, as well as to more exotic magnetic excitations, e.g. antiskyrmions,
bimerons, biskyrmions, skyrmioniums, etc. \cite{Gobel2021}.

\noindent\textsf{\color{blue} Acknowledgments. --- }
We thank A. Fraerman, M. Kuznetsov, and M. Shustin for useful discussions. We are grateful to O. Tretiakov and P. Vorobyev for collaboration on a related project. The work was funded by the Russian Science Foundation under the Grant No. 21-42-04410. The authors gratefully acknowledge the computing time provided to them at computer facilities at Landau Institute.


\begin{thebibliography}{99}

\bibitem{Ryazanov2004} V. V. Ryazanov, V. A. Oboznov, A. S. Prokofiev,
V. V. Bolginov, and A. K. Feofanov, {\it Superconductor--ferromagnet--
superconductor $\pi$-junctions}, J. Low Temp. Phys. {\bf 136}, 385 (2004).

\bibitem{Lyuksyutov2005} I. F. Lyuksyutov and V. L. Pokrovsky, {\it  Ferromagnet--superconductor
hybrids}, Adv. Phys. {\bf 54}, 67 (2005).

\bibitem{Buzdin2005} A. I. Buzdin, {\it Proximity effects in superconductor--ferromagnet
heterostructures}, Rev. Mod. Phys. {\bf 77}, 935 (2005).

\bibitem{Bergeret2005} F. S. Bergeret, A. F. Volkov, and K. B. Efetov,
{\it Odd triplet superconductivity and related phenomena
in superconductor--ferromagnet structures}, Rev. Mod. Phys. {\bf 77}, 1321 (2005).

\bibitem{Eschrig2015} M. Eschrig, {\it Spin-polarized supercurrents for spintronics:
A review of current progress}, Rep. Prog. Phys. {\bf 78}, 104501 (2015).

\bibitem{Back2020} C. Back, V. Cros, H. Ebert, K. Everschor-Sitte, A. Fert,M. Garst, T. Ma, S. Mankovsky, T. L. Monchesky, M. Mostovoy, N. Nagaosa, S. S. P. Parkin, C. Pffeiderer, N. Reyren, A. Rosch, Y. Taguchi, Y. Tokura, K. von Bergmann, and J. Zang, {\it The 2020 skyrmionics roadmap}, J. Phys. D: Applied Phys. {\bf 53}, 363001 (2020).

\bibitem{Gobel2021} B. G\"obel, I. Mertig, and O. A. Tretiakov, {\it Beyond
skyrmions: Review and perspectives of alternative magnetic
quasiparticles}, Phys. Rep. {\bf 895}, 1 (2021).

\bibitem{Zlotnikov} A. O. Zlotnikov, M. S. Shustin, and A. D. Fedoseev, {\it Aspects of topological superconductivity in 2D systems: Noncollinear magnetism, skyrmions, and higher-order topology}, J. Supercond. Nov. Magn. {\bf 34}, 3053 (2021).

\bibitem{Bogdanov1989} A. N. Bogdanov and D. Yablonskii, {\it Thermodynamically
stable ``vortices'' in magnetically ordered crystals.
The mixed state of magnets}, Sov. Phys. JETP {\bf 68}, 101 (1989).

\bibitem{Pershoguba2016} S. S. Pershoguba, S. Nakosai, and A. V. Balatsky,
{\it Skyrmion--induced bound states in a superconductor}, Phys. Rev. B {\bf 94}, 064513 (2016).

\bibitem{Poyhonen2016} K. P\"oyh\"onen, T. Ojanen A.Weststr\"om, S. S. Pershoguba,
and A. V. Balatsky, {\it Skyrmion-induced bound states in a p-wave superconductor}, Phys. Rev. B {\bf 94}, 214509
(2016).

\bibitem{Chen2015} W. Chen and A. P. Schnyder, {\it Majorana edge states
in superconductor-noncollinear magnet interfaces}, Phys.
Rev. B {\bf 92}, 214502 (2015).

\bibitem{Yang2016} G. Yang, P. Stano, J. Klinovaja, and D. Loss, {\it Majorana
bound states in magnetic skyrmions}, Phys. Rev. B {\bf 93},
224505 (2016).

\bibitem{Gungordu2018} U. G\"ung\"ord\"u, S. Sandhoefner, and A. A. Kovalev, {\it Stabilization
and control of majorana bound states with
elongated skyrmions}, Phys. Rev. B {\bf 97}, 115136 (2018).

\bibitem{Mascot2019} E. Mascot, S. Cocklin, S. Rachel, and D. K. Morr, {\it Dimensional
tuning of majorana fermions and real space
counting of the Chern number}, Phys. Rev. B {\bf 100}, 184510 (2019).

\bibitem{Rex2019} S. Rex, I. V. Gornyi, and A. D. Mirlin, {\it Majorana bound
states in magnetic skyrmions imposed onto a superconductor}, Phys. Rev. B {\bf 100}, 064504 (2019).

\bibitem{Garnier2019} M. Garnier, A. Mesaros, and P. Simon, {\it Topological
superconductivity with deformable magnetic skyrmions}, Commun. Phys. {\bf 2}, 126 (2019).

\bibitem{Rex2020} S. Rex, I. V. Gornyi, and A. D. Mirlin, {\it Majorana modes
in emergent-wire phases of helical and cycloidal magnetsuperconductor
hybrids}, Phys. Rev. B {\bf 102}, 224501 (2020). 

\bibitem{Gungordu2022} U. G\"ung\"ord\"u
and A. A. Kovalev, {\it Majorana bound states with chiral magnetic textures}
J. of Appl. Phys. {\bf 132}, 041101 (2022).

\bibitem{Nothhelfer2022} J. Nothhelfer, S. A. D\'iaz, S. Kessler, T. Meng, M. Rizzi, K. M. D. Hals, and K. Everschor-Sitte, {\it Steering Majorana braiding via skyrmion-vortex pairs: A scalable platform}, Phys. Rev. B {\bf 105}, 224509 (2022).

\bibitem{Yokoyama2015} T. Yokoyama and J. Linder, {\it Josephson effect through magnetic skyrmions}, Phys. Rev. B {\bf 92}, 060503(R) (2015).

\bibitem{Proshin2022} V. A. Tumanov, V. E. Zaitseva, Yu. N. Proshin, {\it Critical temperature of superconductor/ferromagnet nanostructure near magnetic skyrmion}, Pis'ma v ZhETF {\bf 106}, 443 (2022). 




\bibitem{Hals2016} K. M. D. Hals, M. Schecter, and M. S. Rudner,
{\it Composite topological excitations in ferromagnet--superconductor
heterostructures}, Phys. Rev. Lett. {\bf 117}, 017001 (2016).

\bibitem{Baumard2019} J. Baumard, J. Cayssol, F. S. Bergeret, and A. Buzdin,
{\it Generation of a superconducting vortex via N\'eel
skyrmions}, Phys. Rev. B {\bf 99}, 014511 (2019).


\bibitem{Dahir2019} S. M. Dahir, A. F. Volkov, and I. M. Eremin, {\it Interaction
of skyrmions and Pearl vortices in superconductor -- chiral
ferromagnet heterostructures}, Phys. Rev. Lett.
{\bf 122}, 097001 (2019).

\bibitem{Menezes2019} R. M. Menezes, J. F. S. Neto, C. C. de Souza Silva,
and M. V. Milo\'sevi\'c, {\it Manipulation of magnetic
skyrmions by superconducting vortices in ferromagnet--superconductor
heterostructures}, Phys. Rev. B {\bf 100},
014431 (2019).

\bibitem{Dahir2020} S. M. Dahir, A. F. Volkov, and I. M. Eremin, {\it Meissner
currents induced by topological magnetic textures
in hybrid superconductor/ferromagnet structures}, Phys.
Rev. B {\bf 102}, 014503 (2020).


\bibitem{Andriyakhina2021} E. S. Andriyakhina and I. S. Burmistrov, {\it Interaction of a N\'eel-type skyrmion with a superconducting vortex}, Phys. Rev. B {\bf 103}, 174519 (2021).


\bibitem{Petrovic2021} A. P. Petrovi\'c, M. Raju, X. Y. Tee, A. Louat, I. Maggio-Aprile,
R. M. Menezes, M. J. Wyszy\'nski, N. K. Duong, M. Reznikov,
Ch. Renner, M. V. Milosevi\'c, and C. Panagopoulos, {\it Skyrmion-
(Anti)Vortex Coupling in a Chiral Magnet-Superconductor
Heterostructure}, Phys. Rev. Lett. {\bf 126}, 117205 (2021).

\bibitem{Pearl1964} J. Pearl, {\it Current distribution in superconducting films carrying
quantized fluxoids}, Appl. Phys. Lett. {\bf 5}, 65 (1964).

\bibitem{AbrikosovBook} A. A. Abrikosov, {\it Fundamentals of the Theory of Metals} (North-Holland, Amsterdam, 1988).


\bibitem{Kawaguchi2016} Y. Kawaguchi, Y. Tanaka, and N. Nagaosa, {\it Skyrmionic
magnetization configurations at chiral magnet/ferromagnet heterostructures},
Phys. Rev. B {\bf 93}, 064416 (2016).

\bibitem{OOMMF} M. J. Donahue and D. G. Porter, {\it OOMMF User's Guide, Version 1.0},  Interagency Report NISTI {\bf 6376} (1999).

\bibitem{Ubermag} M. Beg and M. Lang and H. Fangohr, {\it Ubermag: Towards more effective micromagnetic workflows}, IEEE Transactions on Magnetics {\bf 58}, 1 (2022).


\bibitem{Rohart2013} S. Rohart and A. Thiaville, {\it Skyrmion confinement in ultrathin
film nanostructures in the presence of Dzyaloshinskii--Moriya
interaction}, Phys. Rev. B {\bf 88}, 184422 (2013).

\bibitem{Vadimov2018} V. L. Vadimov, M. V. Sapozhnikov, and A. S. Mel'nikov, {\it Magnetic skyrmions in ferromagnet-superconductor
(f/s) heterostructures}, Appl. Phys. Lett. {\bf 113}, 032402 (2018).

\bibitem{Gonzalez2022} L. Gonz\'alez-G\'omez, J. Castell-Queralt, N. Del-Valle, and C. Navau, {\it Mutual interaction between superconductors and ferromagnetic skyrmionic structures in confined geometries}, Phys. Rev. Applied {\bf 17}, 034069 (2022).

\bibitem{Neto2022} J. F. Neto and C. C. de Souza Silva, {\it Mesoscale phase separation of skyrmion-vortex matter in chiral-magnet-superconductor heterostructures}, Phys. Rev. Lett. {\bf 128}, 057001 (2022).

\end{thebibliography}

\end{document}